\documentclass{PoS}
\title{Central Acceptance Testing for Camera Technologies for the Cherenkov 
Telescope Array}

\ShortTitle{Central Acceptance Testing for Camera Technologies for CTA}

\author{\speaker{A.~Bonardi}$^1$, T.~Buanes$^2$, 
P.~Chadwick$^3$, F.~Dazzi$^4$, A.~F\"orster$^5$, J.~R.~H\"orandel$^1$, 
M.~Punch$^6$, R.~M.~Wagner$^7$ for the CTA Consortium\footnote{Full consortium author list at http://cta-observatory.org}\\
        $^1$Radboud University Nijmegen, Nijmegen (the Netherlands)\\
        $^2$University of Bergen, Bergen (Norway)\\
        $^3$Durham University, Durham (United Kingdom)\\
        $^4$Max Planck Institute for Physics, Munich (Germany)\\
        $^5$Max Planck Institute for Nuclear Physics, Heidelberg (Germany)\\
        $^6$Linnaeus University, V\"axj\"o (Sweden) - CNRS-APC, Paris (France)\\
        $^7$Oskar Klein Centre, Stockholm University, Stockholm (Sweden)\\
        E-mail: \email{a.bonardi@astro.ru.nl}
        }

\abstract{The Cherenkov Telescope Array (CTA) is an international initiative to 
build the next generation ground based very-high energy gamma-ray observatory. 
It will consist of telescopes of three different sizes, employing several different 
technologies for the cameras that detect the Cherenkov light from the observed 
air showers.  In order to ensure the compliance of each camera technology with 
CTA requirements, CTA will perform central acceptance testing of each camera 
technology. To assist with this, the Camera Test Facilities (CTF) work package 
is developing a detailed test program covering the most important performance, 
stability, and durability requirements, including setting up the necessary 
equipment. Performance testing will include a 
wide range of tests like signal amplitude, time resolution, dead-time 
determination, trigger efficiency, performance testing under temperature and 
humidity variations and several others. These tests can be performed on fully-integrated cameras using a portable setup at the camera construction sites. In addition, two different setups for performance tests on camera sub-units are 
being built, which can provide early feedback for camera development. Stability and durability tests will 
include the long-term functionality of movable parts, water tightness of the 
camera housing, temperature and humidity cycling, resistance to vibrations 
during transport or due to possible earthquakes, UV-resistance of materials and 
several others. Some durability tests will need to be contracted out because 
they will need dedicated equipment not currently available within CTA.
The planned test procedures and the current status of the test facilities will 
be presented.
}

\FullConference{The 34$^{th}$ International Cosmic Ray Conference,\\
		30 July - 6 August, 2015\\
		The Hague, The Netherlands}

\begin{document}

\section{Central Acceptance Testing Mission and Goals}

The Cherenkov Telescope Array (CTA) is an international initiative to build the next generation ground based very-high energy gamma-ray observatory. It will consist of two arrays, one in the Northern and one in the Southern Hemisphere, to achieve full-sky coverage. To obtain a wide energy coverage it will use telescopes of three different sizes.\cite{CTA_concept}.\\
Several different technologies 
for the Cherenkov light cameras, which detect airshower-induced Cherenkov light, have been independently developed for the three CTA telescope classes over the last years. These cameras have to face harsh conditions with no protection but their own housing, hence their resistance to environmental influences is a major concern.
CTA has defined a comprehensive set of requirements to ensure stable and high-quality data taking and has put the Camera Test Facilities (CTF) work package in place to perform a central, homogeneous and standardized technology evaluation for all technologies.
CTF is developing a detailed test program to cross-check the compliance with the most important performance, stability, and durability requirements. In addition, CTF is designing and building the necessary equipment for these tests, will perform the systematic testing of all camera technologies, and will provide performance test systems to be installed on the CTA sites.

\section{Camera Test Strategy}

The CTF tests will cover the most important CTA requirements relating to performance, mechanical robustness, and long-term durability. The prototypes of all different camera technologies are required to pass these tests before being accepted for CTA. In addition, CTF is ready to perform preliminary tests on camera sub-units, so that possible problems can be detected and corrected at an earlier stage during the development.\\
Performance tests will be performed by CTF members with CTF equipment. For camera sub-units tests will be performed at CTF test facilities  (see Sect.~\ref{module_test_setup},\ref{cluster_test_setup}); for full cameras with a mobile test setup (see Sect.~\ref{camera_test_setup}) at the respective camera location due to the costs and risks involved in transporting cameras of several meters in size and several tonnes in weight.
For stability and durability testing CTF will offer facilities (see Sect.~\ref{durham_climate_chamber},\ref{durham_salt_mist_chamber}). 
Wind and vibration tests (see Sect.~\ref{mechanical_test}) require large-scale, dedicated facilities, and 
no suitable equipment is available within CTA. It is more cost- and time-effective to contract these tests out.\\

Time and manpower needed for the full set of tests on each camera are summarized in Tab.~\ref{time_requirement}. Several long-running tests are automated and do not need attendance for most of the test duration.
\begin{table}[h!]
\centering
\begin{tabular}[l]{l c}
{\quad \bf Test Category} \qquad & {\bf Time}\\
\hline
\hline
\quad Performance Tests \qquad  & 1 month \qquad \\
\quad Mechanical Tests \qquad  &  1.5 months \qquad \\
\quad Long-term Durability Tests \qquad \qquad &  3 months \qquad \\
\hline
\hline
\end{tabular}
\caption{Time needed for the full set of tests on a single camera.}
\label{time_requirement}
\end{table}

\subsection{Performance Tests}
\label{performance_test}
The CTA performance requirements aim at an improvement of a factor 10 in 
sensitivity compared to current Cherenkov telescope experiments and
at an extension of the energy range covered to both lower and higher
energies. To cross-check if the cameras meet these requirements,  
the performance tests will cover all parameters which can influence 
the camera sensitivity.
 
For the performance tests the following instrumentation is needed:
\begin{itemize}

 \item A {\bf dark room}, 
large enough to accommodate the Cherenkov camera and the test equipment.

 \item A {\bf primary} and a {\bf secondary pulsed light source}, both 
with peak wavelengths between 330 and 430 nm and Gaussian pulse shapes for 
simulating Cherenkov signals. The dynamic range of the output intensity
of the primary light source must be larger than $10^{5}$ and the pulse 
duration adjustable between 1.5 and 5.5 ns.

 \item A {\bf continuous light source} with homogeneous emission and 
a peak wavelength between 500 and 600 nm for simulating the night-sky 
background (NSB). The intensity must be adjustable to achieve  
NSB intensities equivalent to 0.24, 0.5, and 1.2 photons/(ns~sr~cm$^{2}$)
on the focal plane.

 \item A {\bf survival test light source} for simulating accidental 
illumination of the camera focal plane by intensive light. 
Its wavelength distribution must cover most of the visible spectrum and 
the intensity has to be equal to $10^{6}$~photons/(ns~sr~cm$^{2}$)~$\pm~30\%$.

\end{itemize}

The planned performance tests are briefly listed in Tab.~\ref{table_performance_test}.
\begin{table}[h!]
\centering
\begin{tabular}[c]{p{2.88cm} | p{8.35cm} | p{2.63cm}}
{\bf Performance Test} & {\bf Short Description} & {\bf Light Sources}\\
\hline
\hline
{\bf Pixel resolution} & Measurement of ~ the ~ noise ~ level and ~ the ~ charge resolution & Primary with and without NSB\\ 
\hline
{\bf Pixel timing} & Measurement~of~the~time~resolution~and~relative \qquad synchronization & Primary\\
\hline
{\bf Cross-talk} & Measurement of signal contamination to neighboring pixels & Primary\\
\hline
{\bf Trigger \qquad \quad performances} & Determination of trigger threshold, timing, and dead time & Primary with and without NSB\\
\hline
{\bf Event mixing} & Determination of possible mixing of subsequent events in the readout using two different light intensities &  Primary \qquad and secondary\\
\hline
{\bf Power \qquad line variations} & Determination~ of ~robustness~to~power line ~ variations & Primary\\
\hline
{\bf Bright \qquad light exposure} & Measurement ~ of ~ recovery ~ time ~ after ~ accidental exposure to bright light & Survival\\
\hline
\hline
\end{tabular}
\caption{List of performance tests to be performed on each camera technology. NSB indicates the continuous light source: tests with NSB will be performed at the three intensity levels described in the text.}
\label{table_performance_test}
\end{table}
\subsection{Mechanical Tests}
\label{mechanical_test}
CTA requirements define a lifetime of more than 15 years for the cameras
under continuous outdoor operation while at the same time limiting the
allowed maintenance significantly.
The mechanical tests cross-check that all mechanical components of the cameras
satisfy the CTA reliability and lifetime requirements. 
They mainly consist of the following:
\begin{enumerate}
 \item {\bf Test of movable parts:} All movable parts will undergo long-term
testing under different camera orientations.
 \item {\bf Wind load:} In addition, movable parts that need to be 
operational under significant wind loads to protect the camera (e.g. the
lid in front of the focal plane) will be checked for their reliability
under such wind loads.
 \item {\bf Mechanical impact:} To simulate the impact of e.g. 
hailstones, steel balls will be dropped onto different parts of the camera 
housing (all lids closed). Neither the camera body nor any movable part 
must suffer any serious damage.
 \item {\bf Water tightness:} 
Using a regular garden hose, the watertightness of the camera housing and its 
resistance to rain  will be checked.
 \item {\bf Vibration testing:} The cameras need to survive vibration due to 
wind or possible earthquakes and vibrations induced during 
transport, possibly on gravel roads. A test program to simulate
typical vibrations is foreseen.
\end{enumerate}

The water tightness test needs to be performed after most other mechanical
and durability tests to prove that the water tightness has not been affected
by degradation of seals during e.g. temperature cycling, UV or impact testing.

\subsection{Long-Term Durability Tests}
\label{long_term_test}

CTA requirements define a set of environmental parameters, both for performing observations with the telescopes and for their survival when shut down. Typical requirements include a temperature range from 
$-15^{\circ}$~C to $25^{\circ}$~C for operations and from
$-20^{\circ}$~C to $40^{\circ}$~C for survival. Other requirements cover parameters such as 
humidity, wind speeds, and hailstone sizes, among others.
The long-term durability tests aim to check that 
the cameras will meet the lifetime requirement if exposed to 
environmental conditions within the survival requirements.
The tests are briefly summarized in the following:

\begin{enumerate}

 \item {\bf Temperature and humidity cycling:} The camera is placed inside a 
climate chamber and the temperature and the humidity are varied within the 
CTA survival requirements ($-20~^{\circ}\rm{C}<\rm{T}~<~40~^{\circ}\rm{C}$, 
$2\%<\rm{H}<100\%$) for one 1 month. The integrity of camera seals will be 
verified by repeating the water tightness test.

 \item {\bf Solar radiation exposure:} The camera is placed inside a climate chamber with temperature equal to 40~$^{\circ}$C and exposed to radiation of approximately twice the solar intensity. After 2 weeks' exposure, the integrity of camera housing seals will be verified by repeating the rain test.

 \item {\bf Salt fog exposure:} By placing the camera or samples materials employed in the camera (if the camera is too large) in a salt mist chamber at temperature equal to 20~$\pm~5~^{\circ}$C and vaporizing a $5\%$ sodium chloride water solution, the resistance of camera housing to salt fog exposure will be tested. After 2 weeks, any damage to camera body will be investigated by visual inspection and repeating the rain test (in the case of complete camera housings). 
\end{enumerate}

\section{The CTF Test Facilities}
\label{ctf_facilities}
In order to perform all the tests described in Sects.~\ref{performance_test},~\ref{mechanical_test},~and~\ref{long_term_test} both on full camera and on camera sub-units, CTF has made several facilities available. These are described in the following subsections. 

\subsection{The CTF Common Light Source}
\label{common_light_source}

\begin{figure}[!ht]
\centering
\includegraphics[width=.6\textwidth]{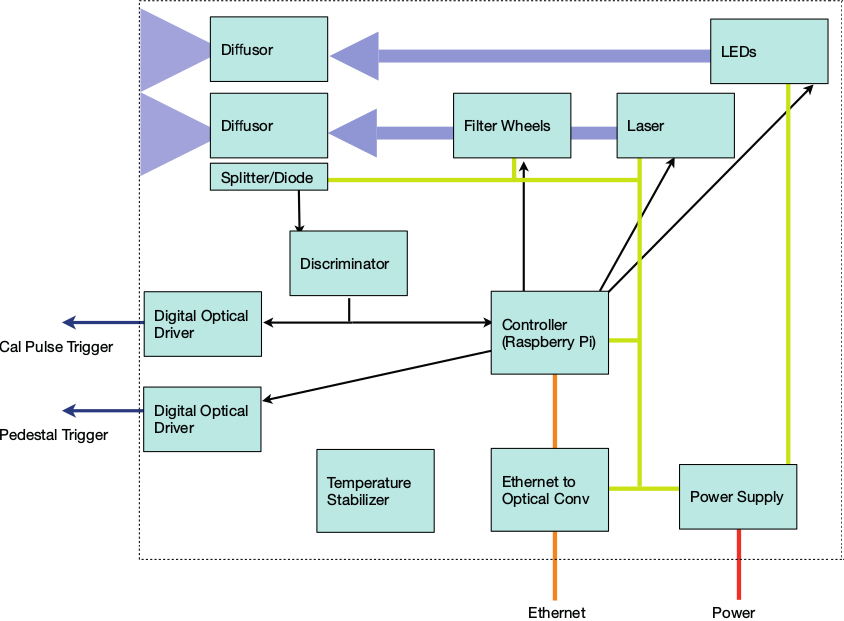}
\caption{Design scheme of the common CTF light source.}
\label{scheme_common_light_source}
\end{figure}

A light source (hereafter ``CTF common light source'') has been developed at Oskar Klein Centre for Cosmoparticle Physics/Stockholm University, to be used in the performance tests described in Sect.~\ref{performance_test}. The CTF common light source will consist of a pulsed laser and a continuous light source based on a LED, both of them housed in a wooden box. The CTF common light source will be used by all the CTF test facilities for performance tests (see Sects.~\ref{module_test_setup},~\ref{cluster_test_setup},~and~\ref{camera_test_setup}) as primary and continuous light source.\\
Two remotely controlled filter wheels will modulate the intensity of the pulsed light source, resulting in 64 achievable intensity levels. To avoid back-reflection of the beam into the light source, the beam splitter and the filter wheels will be slightly misaligned with the beam axis. The pulsed light source will illuminate a beam splitter to couple out a small fraction of the light for an
external trigger of the data acquisition.
Quartz optical fibres and GRIN-type lenses (i.e. lenses focusing light through a refractive index gradient instead of a curved surface) will be used to guide the light from the two light sources to an Ulbricht-type sphere diffuser, so as to physically displace the light sources from the test setup and to be less prone to changing environmental conditions in the test area, e.g. changing temperature or humidity.\\
The scheme of the design of the CTF common light source is shown in fig.~\ref{scheme_common_light_source}.

\subsection{The Module Test Setup with Temperature Control}
\label{module_test_setup}
\begin{figure}[!ht]
\centering
\includegraphics[width=.9\textwidth]{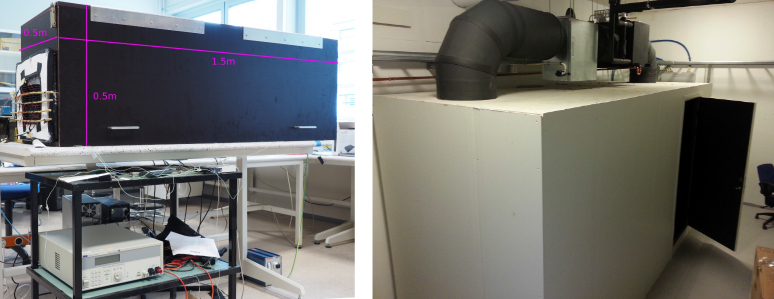}
\caption{\emph{Left}: The Module Test Setup as currently installed at Bergen University. \emph{Right}: The climate dark room ($4\,\rm{m} \times 2\,\rm{m} \times 1.5\,\rm{m}$, $\rm{L \times W \times H}$) for cluster and small camera testing at Linnaeus University in V\"axj\"o.}
\label{Bergen_Vaxjo_setups}
\end{figure}

The test setup at University of Bergen (shown in fig.~\ref{Bergen_Vaxjo_setups}-left) is designed for executing the performance tests (see Sect.\ref{performance_test}) on
modules, i.e., the smallest camera sub-unit with common
electronics. Typically modules consist of $7-64$ camera pixels. 
The setup consists of a temperature-controlled dark box with two reference photo sensors, and 
will use the CTF common light source (see Sect.~\ref{common_light_source}).
The box is 1.5~m $\times$ 0.5~m $\times$
0.5~m, allowing ample room for housing the test specimens. It
has been constructed out of wood so that the box will not act as a
resonator. If electromagnetic shielding proves to be necessary, this will be
fitted on the outside of the box. The inside of the box is clad with
Styrofoam covered with black fabric. One face is equipped with an
aluminium plate, on which 24 Peltier
elements (0.9 kW total cooling power) are placed for temperature control. A set of small fans will
ensure uniform temperature inside the box. The temperature is controlled by a controller
written in LabView, taking input from one or more thermocouples at several positions inside and
outside the box. 
Both the pulsed and the continuous light from the CTF common light source
are diffused to ensure spatial uniformity over the test
specimen. To monitor uniformity in both space and time, the
module test setup is equipped with two PMTs that will be read out
separately from the camera module readout.
Power and data connections for the test specimen are arranged on a
replaceable plate on one short side of the box, allowing them to be
customized to each camera design if necessary.

\subsection{The Cluster Test Setup with Temperature Control}
\label{cluster_test_setup}
The cluster test facility at Linnaeus University in V\"axj\"o (shown in fig.~\ref{Bergen_Vaxjo_setups}-right) is designed to perform performance tests (see Sect.\ref{performance_test}) on groups of modules, known as ``clusters'', which include a part of the back-plane used for communication of trigger signals between modules. The facility can also be used for testing smaller cameras, such as dual mirror telescope cameras.\\
For this purpose, a temperature-controlled dark-room has been constructed, with dimensions of $4 \times 2 \times 1.5\,\rm{m}$ ($\rm{L \times W \times H}$),
constructed essentially from wood and plasterboard (so that the room will not act as a
resonator for electromagnetic waves), placed inside a windowless room.
A computer-controlled air-conditioning  unit has been installed to evacuate the power dissipated from modules or
clusters. The power to be expelled has been estimated for the ensemble of clusters (up to 140 PMTs) or smaller cameras to be below $0.5\,\rm{kW}$
and a further margin of $0.5\,\rm{kW}$ has been allowed for ancillary equipment within the dark-room.
This air-conditioning unit will allow programmed variation of the temperature within the dark-room in a range from $5^\circ$ to $30^\circ$,
which covers the temperature range to be expected within operational cameras.
The CTF common light source (see Sect.~\ref{common_light_source}) will be used as primary and continuous light source, while a second less-flexible light flasher will be used as secondary light source for ``event mixing'' tests (see Tab.~\ref{table_performance_test}).
Masks or light-fibre guides, to perform tests requiring simulated individual Cherenkov images, will be used based on the designs being developed for the Mobile Test Facility.

\subsection{The Mobile Camera Test Setup}
\label{camera_test_setup}
\begin{figure}[!ht]
\centering
\includegraphics[width=1.0\textwidth]{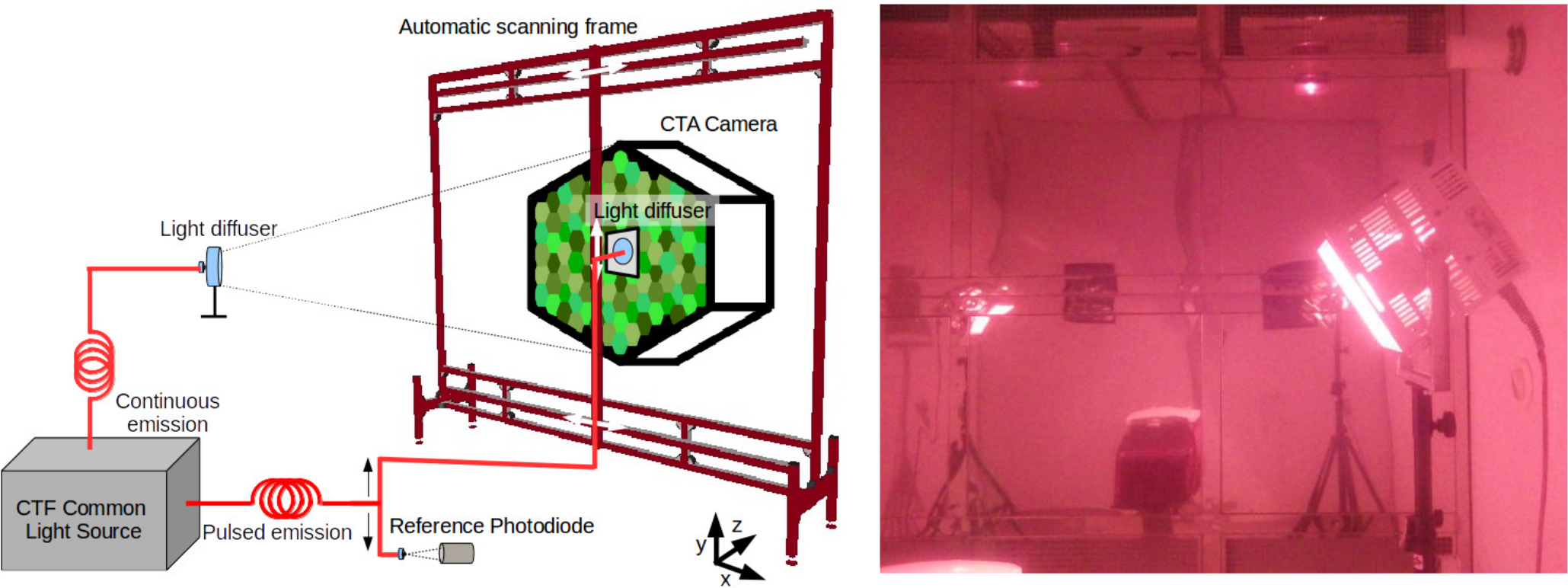}
\caption{\emph{Left}: Scheme of the mobile test setup. For clarity, only the CTF common light source is drawn and the CTA camera is horizontal. \emph{Right}: Solar irradiation test on a CHEC camera housing prototype performed in Durham climate chamber.}
\label{Nijmegen_Durham_setups}
\end{figure}

The mobile test setup has been developed at the Radboud University Nijmegen for executing performance tests on full-equipped CTA cameras at their own location. Since the setup will be moved from site to site, it must be as light weight and flexible in deployment as possible.\\ 
A steering system, consisting of two Bosch Rexroth motorized linear modules mounted on a rectangular frame as support and with strokes equal to 2.950 m and 2.450 m respectively, enable the largest CTA camera to be scanned along both axes with a resolution of $\sim0.1$~mm. The supporting frame will have leveling feet and adjustable height and tilting angle, in order to face the camera focal plane in each possible configuration. The CTF common light source (see Sect.~\ref{common_light_source}) will be used as primary and continuous light source, while a Horiba N-390 NanoLED pulsed LED ($\Delta t = 1.2$~ns, P~=~11~pJ/pulse) will be used as secondary light sources for ``event mixing'' tests (see Tab.~\ref{table_performance_test}). Light from the primary and secondary light sources will be carried to the steering system by modular multimode optical fiber bundles, which will modulate the temporal shape of the pulsed emission.
The light spot shape and size on the camera focal plane will be adjusted by a system of lenses and masks installed on the steering system. The intensity and stability of the light emitted by both pulsed light sources will be monitored by a photodiode. 
A portable PC will directly control the steering system, the light sources and the data analysis, and indirectly the data acquisition, which will be performed by camera electronics. The conceptual scheme of the mobile test setup is shown in fig.~\ref{Nijmegen_Durham_setups}-left.

\subsection{Climate Chamber}
\label{durham_climate_chamber}
The climate chamber at Durham University has a working volume of 3~m~$\times$~3~m~$\times$~3~m. It has been operative for some years and is capable of cooling to $-25~^{\circ}$C and heating to $40~^{\circ}$C, with controlled ramp-up/ramp-down temperature gradients. It can be set to dry air purge to prevent frosting and condensation, but otherwise operates at ambient humidity. In addition, it has a removable roof panel to allow overhead crane access to assist with moving large cameras.\\ 
The climate chamber can also be used for radiation tests, as shown in fig.~\ref{Nijmegen_Durham_setups}-right, thanks to two ATLAS Solar Constant 1200 lamps. These are equipped with metal halide arc lamps which generate a quasi-continuous spectrum very close to the Solar one.

\subsection{Salt Mist Chamber}
\label{durham_salt_mist_chamber}
A salt mist chamber with a volume of 400 l and based on the standard BS 12373-10:1999 has been assembled in Durham. 
The salt mist is pumped into the chamber through an atomizing nozzle, which is fed with from a humidifier that is supplied with salt water solution and air. The air supply has a pressure of around 100 kPa, the level of which is guaranteed by an exhaust vent connected with the outside. The chamber design prevents any condensate dripping from overhead and from the side walls onto the tested item, and the condensate on the chamber floor is removed by a siphon. The whole system is regulated using a stopcock control.

\section{Conclusions}
In order to validate the specifications of all CTA camera technologies, the Camera Test Facilities (CTF) work package is developing a detailed test program as the acceptance procedure for the CTA camera prototypes. CTF is also setting up the equipment for performing those tests, and it will be ready to execute camera full testing before the end of 2015. In the meanwhile, CTF equipment is already available for performance tests on camera sub-units and for environmental tests on fully-equipped cameras.

\acknowledgments
We gratefully acknowledge support from the agencies and organizations 
listed under Funding Agencies at this website: http://www.cta-observatory.org/.

\end{document}